\documentclass[12pt, fleqn]{article}
\usepackage[cp1251]{inputenc}
\usepackage{latexsym,amsfonts,amssymb}
\usepackage{graphicx}

\usepackage{amsbsy}
\usepackage{amsmath}
\usepackage{epsf}

\newtheorem{remark}{Remark}
\newtheorem{theo}{Theorem}
\newcommand{\bt}{\begin{theo}}
\newcommand{\et}{\end{theo}}
\newcommand{\bd}{\begin{displaymath}}
\newcommand{\ed}{\end{displaymath}}

\newcommand{\be} {\begin{equation}}
\newcommand{\ee} {\end{equation}}
\newcommand{\ba} {\begin{array}{l}}
\newcommand{\ea} {\end{array}}

\newcommand{\p} {\partial}

\begin{document}
\begin{center}

\normalsize

 {\Large \bf A hunter-gatherer--farmer population model: Lie
symmetries, exact solutions and their interpretation}\\
 \medskip

 {\bf Roman CHERNIHA~$^{\dag}$\footnote{\small E-mail: r.m.cherniha@gmail.com} and Vasyl' DAVYDOVYCH~$^{\dag}$\footnote{\small  E-mail: davydovych@imath.kiev.ua}} \\
%% \footnote{\small e-mail: cherniha@imath.kiev.ua }\\
\medskip
{\it $^\dag$~Institute of Mathematics,  NAS of Ukraine,\\ 3
Tereshchenkivs'ka Street, 01601 Kyiv, Ukraine}

\renewcommand{\abstractname}{Abstract}
\begin{abstract}
The Lie symmetry classification of the known three-component
reaction-diffusion system modelling  the spread of an initially
localized population of farmers into a
 region occupied by hunter-gatherers is derived. The Lie symmetries obtained for   reducing the system in question to systems of ODEs
 and constructing exact solutions are applied. Several exact solutions of  traveling front type are found, their properties are identified  and  biological interpretation is discussed.
\end{abstract}
\end{center}

\section{Introduction}

In 1952,   A.\,C. Turing published  the remarkable paper
\cite{turing}, in which a revolutionary idea about mechanism of
morphogenesis (the development of structures in an organism during
 its life) has been proposed. From the mathematical point of view Turing's idea immediately leads
 to construction of reaction-diffusion (RD) systems (not single equations!)
 exhibiting so called Turing instability  (see, e.g., Chapter 14.3 in \cite{mu-89}).
  Nowadays nonlinear RD systems are governing equations for  many well-known nonlinear  %@@
second-order models  used to describe  %@@
various processes in biology \cite{britton,  ku-na-ei-16,mu-89, mu-03}, physics \cite{ames,sa-ga-ku-mi-95}, chemistry \cite{aris-75II}, %@@
 ecology \cite{okubo}.

At the present time, one may claim that
 nonlinear RD
systems have been extensively studied by means of different
mathematical methods, including symmetry-based (group-theoretical)
methods  during the last decades. However, the progress is still
insufficient, in particular, Lie symmetries are not completely
described  for many RD systems arising in applications because of
principal and technical difficulties. For example, although  finding
Lie symmetries of
 %the class of RD systems  (\ref{1})
the two-component RD systems  was initiated about 35 years ago
\cite{zulehner-ames}, this Lie symmetry classification problem (the
terminology
 `group classification problem' is also used in this context)
 was finished  only  in the 2000th  in papers
  \cite{ch-king,ch-ki-00-ad,ch-king2,niki-05} (for constant diffusivities) and
   \cite{ ch-king4,ch-king-06,ibrag-94} (for nonconstant diffusivities).

In the case of nonlinear RD systems with the cross-diffusion, the
problem is still open excepting the case when the system in question
involves a constant cross-diffusion only \cite{niki-05}. Notably,
Lie symmetries of some  nonlinear RD systems with
correctly-specified forms of  cross-diffusion arising in real-world
applications were studied in
\cite{cher-davy-muzy-17,cher-wilh-97,se-om-08, to-ro-va-96}.

In contrast to the two-component systems, the multi-component RD
systems (i.e. those consisting of
 three and more equations)  were not widely examined by symmetry-based methods.
 To the best of our knowledge, the most general results for the multi-component
 RD systems (under essential restrictions on the structure of diffusion coefficients)
 were derived in \cite{ch-king-06}. There are also some studies (see, e.g., \cite{che-dav2013})
 devoted to the Lie symmetry search of the multi-component RD systems  involving only arbitrary parameters (i.e. no any arbitrary functions as parameters).
Because, a complete Lie symmetry classification of the general class
of multi-component RD systems
 is extremely difficult problem, it it reasonable to restrict ourselves to some systems arising
 in real world applications.

In this paper, we examine the three-component model introduced in
\cite{ao-sh-shige-96} for describing the spread of an initially
localized population of farmers into a
 region occupied by hunter-gatherers. Under some assumptions clearly indicated
 in \cite{ao-sh-shige-96}, the spread and interaction between  farmers and hunter-gatherers
  can be modeled as a RD process. The corresponding nonlinear RD system
has the form
\begin{equation}\label{1-1}\begin{array}{l} F_t = d_f F_{xx}+r_fF\left(1-(e_1F+e_2C)/K\right),\\ C_t =
d_c C_{xx}+ r_cC\left(1-(e_1F+e_2C)/K\right)+e_1FH+e_2CH,\\ H_t =
d_h H_{xx}+r_hH\left(1-H/L\right)-e_1FH-e_2CH,
\end{array}\end{equation} where $F(t,x), \ C(t,x)$ and $H(t,x)$ are
densities of the three populations of initial farmers, converted
farmers, and hunter-gatherers, respectively. Parameters $d_f, \ d_c$
and $d_h$ are the positive diffusion constants; $r_f, \ r_c$ and
$r_h$ are the intrinsic growth rates of initial farmers, converted
farmers, and hunter-gatherers, respectively; $K$ and $L$ are the
carrying capacities of farmers and hunter-gatherers;  $e_1$ and
$e_2$ are the conversion rates of hunter-gatherers to initial and
converted farmers. Parameters $e_2, \ r_c$ and $r_h$ are assumed to
be nonnegative, while all other parameters are assumed to be
positive. We note that the equalities   $e_1=e_2$ and  $d_f=d_c=d_h$
are assumed  in \cite{ao-sh-shige-96}. In our opinion, it is very
unlikely  that the three populations of initial farmers, converted
farmers, and hunter-gatherers have the same diffusivity in space,
hence their diffusivities  should be assumed arbitrary, i.e. the
equality  $d_f=d_c=d_h$ can take place only in a special case.

The nonlinear RDS (\ref{1-1}) can be simplified using the following
re-scaling  of the variables
\begin{equation}\label{1-1*}F\rightarrow\frac{K}{e_1}u, \
C\rightarrow\frac{KL}{r_f}v, \ H\rightarrow Lw, \
t\rightarrow\frac{1}{r_f}\,t, \
x\rightarrow\sqrt{\frac{1}{r_f}}\,x\end{equation} and introducing
new notation
\[a_1=\frac{e_2L}{r_f}, \ a_2=\frac{r_c}{r_f},
 \ a_3=\frac{r_h}{r_f}, \ a_4=\frac{K}{r_f}\neq0, \
a_5=\frac{e_2KL}{r^2_f}, \ d_f=d_1, \ d_c=d_2, \ d_h=d_3.\]

Re-scaling  (\ref{1-1*}) in symmetry analysis is  called the
equivalence transformation of system (\ref{1-1}).  Transformation
(\ref{1-1*}) reduce system  (\ref{1-1}) to the equivalent form
\begin{equation}\label{1-2}\begin{array}{l}  u_t = d_1 u_{xx}+u(1-u-a_1v),\\  v_t =
d_2 v_{xx}+ a_2v(1-u-a_1v)+uw+a_1vw,\\  w_t = d_3
w_{xx}+a_3w(1-w)-a_4uw-a_5vw.  \end{array}\end{equation} Hereafter
(\ref{1-2}) is called the hunter-gatherer--farmer (HGF) system and
one is the main object of investigation in this paper. We naturally
assume that $a_4\not=0$  (otherwise $K=0$ in (\ref{1-1})) and
$d_1d_2d_3 \not=0$.

The paper is organized as follows. In Section~\ref{sec:2}, the Lie
symmetry classification of the HGF system (\ref{1-2}) is derived. In
Section~\ref{sec:3}, the most important (from applicability point of
view) cases of system (\ref{1-2}) with nontrivial Lie symmetries are
examined. In particular,  nontrivial Lie ans\"{a}tze are derived and
applied for reducing the systems in question to systems of ODEs. The
reduced systems are analyzed in order to construct exact solutions.
In Section~\ref{sec:4}, the traveling fronts (TFs) of the HGF system
(\ref{1-2}) with correctly-specified coefficients are constructed in
explicit forms. The properties of TFs obtained are analysed and some
biological interpretation is presented. Finally, we briefly discuss
the result obtained and present some conclusions in the last
section.

\section{Main theorem} \label{sec:2}

To find Lie invariance  operators, one needs to consider   system
(\ref{1-2}) as the manifold \[{\cal{M}}=\{S_1=0,S_2=0,S_3=0 \},\]
where \begin{eqnarray}  \nonumber  &&
 S_1 \equiv \ d_1u_{xx}-u_t+u(1-u-a_1v),\\ \nonumber  &&
S_2 \equiv \ d_2v_{xx}-v_t+a_2v(1-u-a_1v)+uw+a_1vw, \\ \nonumber &&
S_3 \equiv \ d_3w_{xx}-w_t+a_3w(1-w)-a_4uw-a_5vw,\end{eqnarray} in
the prolonged space of the  variables
 \[t, x, \ u, v, w, \ u_t, v_t,w_t,
u_{x}, v_{x}, w_x, \ u_{xx}, v_{xx}, w_{xx}, u_{xt}, v_{xt}, w_{xt},
u_{tt}, v_{tt}, w_{tt}.\] According to the Lie invariance criterion,
system (\ref{1-2}) is invariant under the Lie group generated by the
infinitesimal operator \begin{eqnarray}\nonumber  && \hskip1cm X = \xi^0 (t, x, u, v, w)\p_{t} + \xi^1 (t, x, u, v, w)\p_{x} +\\  %@@
\nonumber  && \eta^1(t, x, u, v, w)\p_{u}+\eta^2(t, x, u, v,
w)\p_{v}+\eta^3(t, x, u, v, w)\p_{w},
 \end{eqnarray}  %@@
 if the following Lie's invariance conditions are satisfied:
 \begin{equation}\label{2-3} %@@
\mbox{\raisebox{-1.6ex}{$\stackrel{\displaystyle  %@@
X}{\scriptstyle 2}$}} (S_1)  %@@
 %@@
\Big\vert_{\cal{M}}=0, \quad
\mbox{\raisebox{-1.6ex}{$\stackrel{\displaystyle  %@@
X}{\scriptstyle 2}$}} (S_2)  %@@
 \Big\vert_{\cal{M}}=0, \quad
\mbox{\raisebox{-1.6ex}{$\stackrel{\displaystyle  %@@
X}{\scriptstyle 2}$}} (S_3)  %@@
 \Big\vert_{\cal{M}}=0,
\end{equation}  %@@
where the operator $ \mbox{\raisebox{-1.6ex}{$\stackrel{\displaystyle  %@@
X}{\scriptstyle 2}$}} $  %@@
is the second  %@@
 prolongation of the operator $X$ (see, e.g., \cite{arrigo15,bl-anco-10,fss,olv,ovs}).

  Obviously, system (\ref{1-2})
admits the Lie algebra with the  basic operators
\begin{equation}\label{2-2} P_t=\p_t, \ P_x=\p_x
\end{equation}
because one is invariant with respect to the time and space
translations. It can be easily shown that    (\ref{2-2})  is the
principal
 (trivial) algebra  of system  (\ref{1-2}), i.e.
this is maximal invariance algebra of this system with arbitrary
coefficients $a_j$ and $d_k$. To find all possible extensions of
principal  algebra in the case of the system (\ref{1-2}), one needs
to apply the invariance criterion (\ref{2-3})  and to solve the
corresponding  system of determining equations (DEs). Omitting
rather standard calculations (nowadays they can be done using Maple,
Mathematica etc.), we present the DE system  obtained:
\begin{eqnarray} && \label{2-6} \xi^0_{x}=\xi^0_{u}=\xi^0_{v}
=\xi^0_{w}=\xi^1_{u}=\xi^1_{v}=\xi^1_{w}=0,  \\ && \label{2-7}
\eta^k_{uu}=\eta^k_{uv}=\eta^k_{vv}=\eta^k_{ww}=\eta^k_{uw}=\eta^k_{vw}=0,
\ k=1,2,3,  \\ && \label{2-8}
\eta^1_{xv}=\eta^1_{xw}=\eta^2_{xu}=\eta^2_{xw}=\eta^3_{xu}=\eta^3_{xv}=0,
\\&& \nonumber
(d_1-d_2)\eta^1_{v}=(d_1-d_3)\eta^1_{w}=(d_1-d_2)\eta^2_u=\\
&& \label{2-9} (d_2-d_3)\eta^2_{w}=(d_1-d_3)\eta^3_u=
(d_2-d_3)\eta^3_{v}=0,   \\ && \label{2-10} 2\xi^1_x-\xi^0_t =0,   \
2d_1\eta^1_{xu}+\xi^1_{t}=0, \ 2d_2\eta^2_{xv}+\xi^1_t=0,
  \ 2d_3\eta^3_{xw}+\xi^1_t=0,   \\ && \label{2-11}
\eta^1C^1_u+\eta^2C^1_v+\eta^3C^1_w+(2\xi^1_x-\eta^1_u)C^1=
\eta^1_t-d_1\eta^1_{xx}+\frac{d_1}{d_2}\,\eta^1_vC^2+\frac{d_1}{d_3}\,\eta^1_wC^3,
\\ && \label{2-12} \eta^1C^2_u+\eta^2C^2_v+\eta^3C^2_w+(2\xi^1_x-\eta^2_v)C^2=
\eta^2_t-d_2\eta^2_{xx}+\frac{d_2}{d_1}\,\eta^2_uC^1+
\frac{d_2}{d_3}\,\eta^2_wC^3,   \\ && \label{2-13}
\eta^1C^3_u+\eta^2C^3_v+\eta^3C^3_w+(2\xi^1_x-\eta^3_w)C^3=
\eta^3_t-d_3\eta^3_{xx}+\frac{d_3}{d_1}\,\eta^3_uC^1+\frac{d_3}{d_2}\,\eta^3_vC^2,
\end{eqnarray} where \begin{equation}\label{2-5}\begin{array}{l}
 C^1=u(1-u-a_1v),\\ C^2=a_2v(1-u-a_1v)+uw+a_1vw,\\
 C^3=a_3w(1-w)-a_4uw-a_5vw.\end{array}\end{equation}

 Now we want to find all possible values of the coefficients $a_j$ and $d_k$  leading to extensions of the principal algebra (\ref{2-2}). It means that all inequivalent solutions of the system of DEs (\ref{2-6})--(\ref{2-13}) should be constructed. As a result, the following statement was proved.

\bt\label{th-1} The HGF ystem (\ref{1-2}) with $a_4d_1d_2d_3 \not=0$
   admits a nontrivial Lie algebra of  symmetries if and only if   one  and the
corresponding  symmetry operators  have the forms listed in
Table~\ref{tab1}.
  \et

  \begin{table}
\caption{Lie symmetry operators of the HGF  system
(\ref{1-2})}\medskip
\label{tab1}       % Give a unique label
\begin{tabular}{p{0.2cm}p{5cm}p{2.8cm}p{5cm}}
\hline\hline\noalign{\smallskip} & Reaction terms &Restrictions & Lie symmetries  \\  \hline &&&\\
 1 &
$u(1-u)$  \newline{$a_2v(1-u)+uw$}
\newline{$-a_4uw$}
 & $a_2\neq0$& $\p_t, \ \p_x,\ I=v\p_v+w\p_w$  \\ \hline &&&\\
 2 &
$u(1-u)$  \newline{$uw$}
\newline{$a_3w(1-w)-a_4uw$}
 & $a_3\neq0$& $\p_t, \ \p_x,$ \newline{$X^\infty=P(t,x)\p_v,$} \newline{$P_t=d_2P_{xx}$}  \\ \hline &&&\\
 3 &
$u(1-u)$  \newline{$uw$}
\newline{$-a_4uw$}
 & & $\p_t, \ \p_x,\ I, \ X^\infty$  \\ \hline &&&\\
4 & $u(1-u-a_1v)$ \newline{$v(1-u-a_1v)+uw+a_1vw$}
\newline{$a_3w(1-w)-a_4uw-a_1a_4vw$}  & $d_1=d_2$ \newline{$a_1\neq0$} &
$\p_t, \ \p_x,$ \newline{$Q_1=-a_1u\p_u+u\p_v$}\\ \hline &&&\\
5 & $u(1-u)$ \newline{$v(1-u)+uw$}
\newline{$a_3w(1-w)-a_4uw$}  & $d_1=d_2$ \newline{$a_3\neq0$} &$\p_t, \ \p_x,$ \newline{$u\p_v, \ Q_2=e^t(u-1)\p_v$}\\ \hline &&&\\
6 & $u(1-u)$ \newline{$v(1-u)+uw$}
\newline{$-a_4uw$}  & $d_1=d_2$  & $\p_t, \ \p_x,\ u\p_v, \ I, \
Q_2$
\\
 \hline &&&\\
7 & $u(1-u)$ \newline{$a_4v(1-u)+uw$}
\newline{$-a_4uw$}  & $d_2=d_3$  & $\p_t, \ \p_x,\ e^{a_4t}w\p_v, \
I$
\\ \hline &&&\\
8 & $u(1-u)$ \newline{$uw$}
\newline{$-a_4uw$}  & $d_2=d_3$  & $\p_t, \ \p_x,$ \newline{$w\p_v-a_4w\p_w,\ I, \
X^\infty$}
\\ \hline &&&\\
9 & $u(1-u-a_1v)$ \newline{$v(1-u-a_1v)+uw+a_1vw$}
\newline{$-a_4uw-a_1a_4vw$}  & $d_1=d_2=d_3$ \newline{$a_1\neq0$} & $\p_t, \ \p_x, \ Q_1,$ \newline{$e^t\left(\frac{a_4-1}{a_1}\,u+
(a_4-1)v+\right.$}\newline{$\left.w+\frac{1-a_4}{a_1}\right)\left(\p_u-\frac{1}{a_1}\p_v\right)$}
\\  \hline\hline
\end{tabular}
\end{table}
\begin{table} \centerline{Continuation of Table~\ref{tab1}}\medskip
\begin{tabular}{p{0.2cm}p{5cm}p{2.8cm}p{5cm}}
\hline\hline\noalign{\smallskip} & Reaction terms &Restrictions &  Lie symmetries  \\
\hline&&&\\
 10 &
$u(1-u)$  \newline{$a_2v(1-u)+uw$}
\newline{$-uw$}
 & $d_1=d_2=d_3$ \newline{$a_2\neq0, \ a_2\neq1$}& $\p_t, \ \p_x, \ I,$ \newline{$u\p_v+(a_2-1)(u-1)\p_w$}  \\\hline &&&\\
 11 &
$u(1-u)$  \newline{$v(1-u)+uw$}
\newline{$-uw$}
 & $d_1=d_2=d_3$& $\p_t, \ \p_x,$ \newline{$u\p_v, \ we^t\p_v, \ I, \ Q_2$}  \\ \hline &&&\\
12 & $u(1-u)$ \newline{$uw$}
\newline{$-uw$}  & $d_1=d_2=d_3$  & $\p_t, \ \p_x, \ w\p_v-w\p_w,$ \newline{$u\p_v+(1-u)\p_w,$}  \newline{$
ue^{-t}(\p_v-\p_w),\ I, \  X^\infty$}
\\  \hline\hline
\end{tabular}
\end{table}

\textbf{Sketch of the proof.} In order  to prove the theorem, one
needs to solve  the system of DEs (\ref{2-6})--(\ref{2-13}) with the
functions $C^k \ (k=1,2,3)$  from (\ref{2-5}). Although this is a
standard routine, all possible  special cases (not some of them !)
should be identified and examined in order to obtain a full Lie
symmetry classification.

 It can be noted that
the forms of the functions $\xi^0, \ \xi^1$ and $\eta^k \ (k=1,2,3)$
can be defined independently on the functions $C^k$. In fact,
equations (\ref{2-6})--(\ref{2-8}) can be easily
 integrated:  \begin{eqnarray} \nonumber &&
\xi^0=\xi^0(t), \ \xi^1=\xi^1(t,x), \\ \nonumber &&
\eta^1=r^1(t,x)u+q^1(t)v+h^1(t)w+p^1(t,x), \\ \nonumber &&
\eta^2=r^2(t,x)v+q^2(t)u+h^2(t)w+p^2(t,x), \\ \nonumber &&
\eta^3=r^3(t,x)w+q^3(t)u+h^3(t)v+p^3(t,x),
\end{eqnarray} where $\xi^0, \ \xi^1, \ r^k, \ q^k, \ h^k$ and $p^k$ ($k=1,2,3$) are to-be-determined
functions.

 Now we analyse equations (\ref{2-9}). It  turns out that five different
      cases should be examined depending on diffusion coefficients, namely:
 {\it (I)}  $d_k$  are arbitrary positive constants,
 {\it (II)} $d_1=d_2,$
  {\it (III)} $d_1=d_3,$
 {\it (IV)} $d_2=d_3$ and
 {\it (V)} $d_1=d_2=d_3.$

Let us examine   case {\it (I)}. Because the diffusivities $d_k \
(k=1,2,3)$ are arbitrary constants,  equations (\ref{2-9})
immediately produce $q^k=h^k=0, \ k=1,2,3.$  Equations
(\ref{2-11})--(\ref{2-13}) can be split with respect to the
variables $u, v, w$ and their products $uv,uw, vw, u^2,$ $v^2, w^2$.
As a result, the system of DEs (\ref{2-6})--(\ref{2-13}) reduces to
the form \begin{eqnarray} && \label{2-16} a_1 p^1=0, \ p^1+a_1
p^2=0, \ -a_2 p^2+p^3=0, \ a_4p^3=0, \ a_5p^3=0,\ r^2=r^3,
\\&& \label{2-17} 2\xi^1_x-\xi^0_t =0, r^1+2\xi^1_x=0,  2d_1r^1_{x}+\xi^1_{t}=0,
2d_2r^2_{x}+\xi^1_t=0,
  \ 2d_3r^3_{x}+\xi^1_t=0,  \\ && \label{2-18}
  a_1(r^2+2\xi^1_x)=0, \ a_3(r^2+2\xi^1_x)=0, \ a_5(r^2+2\xi^1_x)=0,
\\ && \label{2-19}d_1 r^1_{xx}-r^1_t+2\xi^1_x -2p^1-a_1 p^2=0,   \\ && \label{2-20}
d_2 r^2_{xx}-r^2_t+2a_2\xi^1_x -a_2p^1-2a_1a_2 p^2+a_1p^3=0,\\ &&
\label{2-21} d_3 r^3_{xx}-r^3_t+2a_3\xi^1_x -a_4p^1-a_5
p^2-2a_3p^3=0,\\ && \label{2-22}d_1 p^1_{xx}-p^1_t+p^1=0, \ d_2
p^2_{xx}-p^2_t+a_2p^2=0, \ d_3 p^3_{xx}-p^3_t+a_3p^3=0.
\end{eqnarray}

Because (\ref{2-16}) is the set of algebraic equations, we find
$p^1=p^3=0$ and $a_1a_2p^2=0,$ while the overdetermined system
(\ref{2-17}) leads to
\[\xi^0_{tt}=\xi^1_{xx}=\xi^1_{t}=r^1_x=r^1_t=r^2_x=0.\]
Hence, equation (\ref{2-19}) produces $\xi^1_x=0$. Having
$\xi^1_x=0$, equations (\ref{2-18}) give $r^2=r^3=0$ provided
$a_1^2+a_3^2+a_5^2\neq0.$ In this case, one can find nontrivial Lie
symmetry only under the restriction  $p^2\neq0$,  hence
$a_1=a_2=a_5=0.$ Thus, the general solution of
(\ref{2-16})--(\ref{2-22}) has the form
\[ \xi^0=c_0, \ \xi^1=c_1, \ p^1=p^3=0,\  r^1=r^2=r^3=0, \ p^2=P(t,x)\]
(hereafter $c_k \ (k=0,1,\dots)$ is arbitrary constant, while the
function $P(t,x)$ is an arbitrary solution of equation
$P_t=d_2P_{xx}$), therefore Case 2 of Table~\ref{tab1} is obtained.

In the case $a_1=a_3=a_5=0$ we  obtain  Cases 1 and 3 of
Table~\ref{tab1}. Thus, case {\it(I)} is completely examined.

Now we turn to case {\it (II)}. Having done a   preliminary
analysis, we  find and \begin{eqnarray} &&
\label{2-23} q^1=q^3=h^k=0, \ k=1,2,3,\\
&& \nonumber p^1=p^3=0, \  a_1 p^2=0 \end{eqnarray} and  derive
%(such as for equations (\ref{2-16}))
the system of DEs \begin{eqnarray} && \label{2-25} 2\xi^1_x-\xi^0_t
=0,   \ 2d\eta^1_{xu}+\xi^1_{t}=0, \ 2d\eta^2_{xv}+\xi^1_t=0,
  \ 2d_3\eta^3_{xw}+\xi^1_t=0,   \\
&& \label{2-26} d r^1_{xx}-r^1_t+2\xi^1_x=0,   d
r^2_{xx}-r^2_t+2a_2\xi^1_x=0,
d_3 r^3_{xx}-r^3_t+2a_3\xi^1_x-a_5 p^2=0, \\
&& \label{2-27}
  a_1(r^2+2\xi^1_x)=0,  \ a_5(r^2+2\xi^1_x)=0,\ a_1(r^3+2\xi^1_x)=0, \ a_3(r^3+2\xi^1_x)=0,
\\
&& \label{2-28}  \left(-1+a_2\right) q^2=0,\
a_1 \left(-1+2 a_2\right) q^2+a_2 \left(r^1+2 \xi^1_x\right)=0,  \\
&& \label{2-29} a_5 q^2+a_4 \left(r^1+2 \xi^1_x\right)=0,  \ a_1
q^2+r^1+2\xi^1_x=0,\\ && \label{2-30}
 a_1 q^2+r^1-r^2+r^3+2 \xi^1_x=0, \\
&& \label{2-31} d q^2_{xx}-q^2_t+\left(a_2-1\right)q^2-a_2 p^2=0, \\
&& \label{2-32} d p^2_{xx}-p^2_t+a_2p^2=0  \end{eqnarray} for
finding all other functions.

It can be seen from  (\ref{2-23}) that new nontrivial Lie symmetries
can exist only if $q^2\neq0$ (otherwise one obtains the result of
case {\it (I)}). Thus, the first equation of (\ref{2-28})
immediately produces $a_2=1,$ while restriction $a_5=a_1a_4$ follows
from  the compatibility condition of equations~(\ref{2-29}).

The further analysis of the system of DEs (\ref{2-25})--(\ref{2-32})
depends on the value of constant $a_1$.

If $a_1\neq0$  then $p^2=0$ and $r^2=r^3=-2\xi^1_x$. As a result,
equations (\ref{2-25}) and (\ref{2-26}) produce $r^2=r^3=0, \
r^1=const,\ \xi^0=c_0,\ \xi^1=c_1.$ The last unknown function $q^2$
can be found from (\ref{2-29}). Hence,
\[\xi^0=c_0, \ \xi^1=c_1, \ \eta^1=-a_1c_3u, \ \eta^2=c_3u, \ \eta^3=0,\]
and  Case 4 of Table~\ref{tab1} is obtained. In a quite similar way,
one examines the subcase $a_1=0$ and arrives at Cases 5--6 of
Table~\ref{tab1}.

The examination of the system of DEs in case  {\it (III)} does not
lead to new system  of the form (\ref{1-2}) with nontrivial Lie
symmetries.

Analysis of case {\it (IV)} leads to systems and Lie symmetries
listed in Cases 7 and 8 of Table~\ref{tab1}, while case  {\it (V)}
produces Cases 9--12 of Table~\ref{tab1}. The relevant calculations
are omitted here.

The proof is now completed. \hfill $\square$

\section{Reduction of the HGF system to ODE systems} \label{sec:3}

In this section, we present examples of reductions of the HGF system
(\ref{1-2}) to ODE systems using the Lie symmetries obtained. If one
compares  system (\ref{1-2}) with the reaction terms arising in
Table~\ref{tab1} with its general form (\ref{1-2}) then are realizes
that Cases 4, 5 and 9 are the most interesting from the
applicability point of view. In fact, all the other cases of
Table~\ref{tab1} lead to the systems of the form (\ref{1-2}) with
too many zero coefficients, hence it is unlikely that such systems
can describe adequately the spread and interaction between  farmers
and hunter-gatherers.
 For this reason, we consider the systems
from Cases 4, 5 and 9 of Table~\ref{tab1} and the relevant linear
combinations of the  Lie symmetries involving  nontrivial operators.
The case of the Lie symmetry operators leading to  plane wave
solutions, especially  TFs,  is examined separately  in
Section~\ref{sec:4}.

First of all, we note that one diffusivity, e.g. $d_1$, can be set
$1$ in (\ref{1-2}) without losing a generality, hence the system
from Case 4 of Table~\ref{tab1} have the form
\begin{equation}\label{3-1}\begin{array}{l} u_t = u_{xx}+u(1-u-a_1v),\ a_1\neq0,\\ v_t =
 v_{xx}+v(1-u-a_1v)+uw+a_1vw, \\ w_t = d w_{xx}+a_3w(1-w)-a_4uw-a_1a_4vw. \end{array}\end{equation}
Let us consider two essentially different  linear combinations of
the Lie symmetry operators of system (\ref{3-1})
\begin{equation}\label{3-2} X=\p_t+\alpha\p_x-\beta a_1u\p_u+\beta u
\p_v \end{equation} and
\begin{equation}\label{3-2*} X=\p_x-\beta a_1u\p_u+\beta u \p_v. \end{equation}
 Hereafter $\alpha$
and $\beta\neq0$ are arbitrary constants.

 Solving the
characteristic equation
\[\frac{dt}{1}=\frac{dx}{\alpha}=\frac{du}{-\beta a_1u}=\frac{dv}{\beta u}\]
corresponding to
 operator (\ref{3-2}) one obtains  the ansatz
\begin{equation}\label{3-3}\begin{array}{l}  u=e^{-\beta a_1t}U(\omega), \ \omega=x-\alpha
t, \\  v=V(\omega)-\frac{1}{a_1}\,e^{-\beta a_1t}U(\omega),\\
w=W(\omega),
\end{array}\end{equation} where $U, \ V$ and $W$ are new unknown
functions. Substituting ansatz (\ref{3-3}) into (\ref{3-1}), we
arrive at the system of ODEs
\begin{equation}\label{3-4}\begin{array}{l} U''+\alpha U'+
U\left(1+a_1\beta-a_1V\right)=0, \\ V''+\alpha V'+
V\left(1-a_1V+a_1W\right)=0,\\ dW''+\alpha W'+
a_3W\left(1-W\right)-a_1a_4VW=0.
\end{array}\end{equation}
One sees that the reduced  system (\ref{3-4}) is nonlinear and the
problem of constructing its exact solutions is still a difficult
task. However, we were able to note the three special cases
\[\mbox{(i)}\ d=a_3=1, \quad \mbox{(ii)} \ d=1, \ a_3=0, \quad \mbox{(iii)} \ d=1, \ a_4=1+a_1+a_3, \ a_3\neq0,\] when the functions $V$ and $W$ can be found, while $U$
satisfies a separate ODE. In fact, if one assumes that the
components $V$  and $W$  have the same structure as the well-known
solution of the Fisher equation \cite{abl-zep} (see  formula
(\ref{3-22}) below) then the cases (i), (ii) and (iii)  lead
 to the exact solutions
\begin{eqnarray}\nonumber &&  u=e^{-\beta a_1t}U(\omega), \ \omega=x-
\frac{5}{\sqrt{6}}\,t, \\ \nonumber &&
v=\frac{1+a_1}{4a_1(1+a_1a_4)}\left(1-
\tanh\left[\frac{1}{2\sqrt{6}}\,\omega\right]\right)^2-\frac{1}{a_1}\,e^{-\beta
a_1t}U(\omega),\\ \nonumber &&w=\frac{1-a_4}{4(1+a_1a_4)}\left(1-
\tanh\left[\frac{1}{2\sqrt{6}}\,\omega\right]\right)^2,
\end{eqnarray}
\begin{eqnarray}\nonumber &&   u=e^{-\beta a_1t}U(\omega), \ \omega=x-
\frac{5\sqrt{a_4}}{\sqrt{6}}\,t, \\ \nonumber &&
v=\frac{1}{4a_1}\left(1-
\tanh\left[\frac{\sqrt{a_4}}{2\sqrt{6}}\,\omega\right]\right)^2-\frac{1}{a_1}\,e^{-\beta
a_1t}U(\omega),\\ \nonumber &&
w=\frac{1-a_4}{4a_1}\left(\left(1-\tanh
\frac{\sqrt{a_4}}{2\sqrt{6}}\,\omega\right)^2-4\right)
\end{eqnarray} and
\begin{eqnarray}\nonumber &&   u=e^{-\beta a_1t}U(\omega), \ \omega=x-
\frac{5\sqrt{1+a_1}}{\sqrt{6}}\,t, \\ \nonumber &&
v=\frac{1}{4a_1}\left(1-
\tanh\left[\frac{\sqrt{1+a_1}}{2\sqrt{6}}\,\omega\right]\right)^2-\frac{1}{a_1}\,e^{-\beta
a_1t}U(\omega),\\ \nonumber && w=1-\frac{1}{4}\left(1-\tanh
\frac{\sqrt{1+a_1}}{2\sqrt{6}}\,\omega\right)^2,
\end{eqnarray}
% of system(\ref{3-1}),
respectively. Here the function $U$ is an arbitrary solution of the
linear ODE \begin{equation}\label{3-25} U''+\alpha U'+
U\left(1+a_1\beta-\kappa_1\left(1-
\tanh\left[\frac{\kappa_2}{2\sqrt{6}}\,\omega\right]\right)^2\right)=0,\end{equation}
where
\[ \kappa_1= \left\{ \begin{array}{l}
%\begin{array}{cc}
\frac{1+a_1}{4(1+a_1a_4)}\hskip0.4cm  \mbox{in case (i)},\\
\frac{1}{4} \hskip1.6cm \mbox{in case (ii)}, \\
\frac{1}{4} \hskip1.6cm \mbox{in case (iii)},
\end{array} \right. \quad \kappa_2= \left\{ \begin{array}{l}1 \hskip1.6cm  \mbox{in case (i)}, \\
\sqrt{a_4} \hskip1.1cm \mbox{in case (ii)},
   \\
\sqrt{1+a_1} \hskip0.5cm \mbox{in case (iii)}.  \end{array} \right.
\]

Ansatz corresponding to operator (\ref{3-2*}) and the reduced system
for system (\ref{3-1}) have the forms
\begin{equation}\label{3-36}\begin{array}{l}  u=U(t)e^{-\beta a_1x},
\quad v=V(t)-\frac{e^{-\beta a_1x}}{a_1}\,U(t),\quad w=W(t)
\end{array}\end{equation} and
\begin{equation}\label{3-37}\begin{array}{l}U'+U(a_1V-1-\beta^2a_1^2)=0,\\
V'+V(a_1V-a_1W-1)=0,\\
W'+W(a_3W+a_1a_4V-a_3)=0.
 \end{array}\end{equation}
We have solved system (\ref{3-37}) assuming that the functions $V$
and $W$ are linearly dependent. In a such way three different cases
\[\mbox{(i)}\ a_3=1, \quad \mbox{(ii)} \ a_3=0, \quad \mbox{(iii)} \ a_4=1+a_1+a_3, \ a_3\neq0\]
hold. Thus the cases (i), (ii) and (iii)  lead
 to the exact solutions
\begin{equation}\label{3-38}\begin{array}{l}
U=\frac{\delta_2e^{(1+\beta^2a_1^2)t}}{\left(1-\delta_1+\delta_1e^t\right)^{\frac{1+a_1}{1+a_1a_4}}},\quad
V=\frac{1+a_1}{a_1(1+a_1a_4)}\frac{\delta_1e^t}{1-\delta_1+\delta_1e^t},\quad
W=\frac{1-a_4}{1+a_1a_4}\frac{\delta_1e^t}{1-\delta_1+\delta_1e^t},
 \end{array}\end{equation}
\[
U=\frac{\delta_2e^{(1+\beta^2a_1^2)t}}{\left(1-\delta_1+\delta_1e^{a_4t}\right)^{\frac{1}{a_4}}},\
V=\frac{\delta_1e^{a_4t}}{a_1\left(1-\delta_1+\delta_1e^{a_4t}\right)},\
W=\frac{(1-a_4)(\delta_1-1)}{a_1\left(1-\delta_1+\delta_1e^{a_4t}\right)}
 \] and
\[
U=\frac{\delta_2e^{(1+\beta^2a_1^2)t}}{\left(1-\delta_1+\delta_1e^{(1+a_1)t}\right)^{\frac{1}{1+a_1}}},\
V=\frac{\delta_1e^{(1+a_1)t}}{a_1\left(1-\delta_1+\delta_1e^{(1+a_1)t}\right)},\
W=\frac{1-\delta_1}{1-\delta_1+\delta_1e^{(1+a_1)t}}
 \]
 (here $\delta_1$ and $\delta_2$ are arbitrary positive
 constants) of system (\ref{3-37}), respectively.

 Let us consider the most
interesting solution (\ref{3-38}) from the applicability point of
view in detail. Substituting (\ref{3-38}) into ansatz (\ref{3-36}),
the three-parameter family of  exact solutions
\begin{equation}\label{3-41}\begin{array}{l}  u=
\frac{\delta_2\exp\left((1+\beta^2a_1^2)t-\beta
a_1x\right)}{\left(1-\delta_1+\delta_1e^t\right)^{\frac{1+a_1}{1+a_1a_4}}},
\\ v=\frac{1+a_1}{a_1(1+a_1a_4)}\frac{\delta_1e^t}{1-\delta_1+\delta_1e^t}-
\frac{\delta_2\exp\left((1+\beta^2a_1^2)t-\beta
a_1x\right)}{a_1\left(1-\delta_1+\delta_1e^t\right)^{\frac{1+a_1}{1+a_1a_4}}},\\
w=\frac{1-a_4}{1+a_1a_4}\frac{\delta_1e^t}{1-\delta_1+\delta_1e^t}
\end{array}\end{equation} of system (\ref{3-1}) with $a_3=1$ is
obtained.

It can be noted that the components of  exact solutions of the form
(\ref{3-41}) are nonnegative on the space interval $ x \in
(0,+\infty)$ provided  the restrictions
\begin{equation}\label{3-41*} \beta=\sqrt{\frac{1-a_4}{a_1(1+a_1a_4)}},
\  a_4<1, \ \delta_1>1, \ \delta_2<\frac{1+a_1}{1+a_1a_4}
\end{equation}
hold. In this case, the solutions possess the asymptotical behaviour
\begin{equation}\label{3-41**}
u \rightarrow \delta_2\delta_1^{-\frac{1+a_1}{1+a_1a_4}}e^{-\beta
a_1x}, \ v \rightarrow
\frac{1+a_1}{a_1(1+a_1a_4)}-\frac{\delta_2}{a_1}\,\delta_1^{-\frac{1+a_1}{1+a_1a_4}}e^{-\beta
a_1x}, \ w \rightarrow \frac{1-a_4}{1+a_1a_4}
\end{equation}
as $t \rightarrow \infty.$

Such behaviour predicts the scenario when the populations of initial
farmers, converted farmers, and hunter-gatherers coexist in
space-time, moreover the distribution of two populations is
inhomogeneous as $t \rightarrow \infty$. Notably, this scenario
occurs at any semi-finite interval (instead of the fixed interval
$(0,+\infty)$) because system (\ref{3-1}) is invariant with respect
to the space translations.

The system and the most general linear combinations of the  Lie
symmetries from Case~5 of Table~\ref{tab1} have the forms
\begin{equation}\label{3-20}\begin{array}{l} u_t = u_{xx}+u(1-u),\\ v_t =
 v_{xx}+v(1-u)+uw, \\ w_t = d w_{xx}+a_3w(1-w)-a_4uw \end{array}\end{equation}
and  \begin{eqnarray} \label{3-21} & & X=\p_t+\alpha\p_x+\beta
u\p_v+\gamma e^t(u-1)\p_v,\\ & & \label{3-42}
 X=\p_x+\beta u\p_v+\gamma e^t(u-1)\p_v.\end{eqnarray}

As one can see, the first equation of system (\ref{3-20}) is the
famous Fisher equation \cite{fi-37} that is not integrable. There
were many attempts to construct its exact solutions
 taking into account some reasonable  initial and  boundary
 conditions. In particular,  the  appropriate  exact  solution in the form of the
 TF
\begin{equation}\label{3-22} u\equiv U(\omega) =
\frac{1}{4}\left(1-\tanh\left[\frac{1}{2\sqrt{6}}\,\omega\right]
\right)^2, \quad \omega=x- \frac{5}{\sqrt{6}}\,t \end{equation}
   was found in
 \cite{abl-zep}. We remind the reader that  a plane wave solution of a PDE, which is nonnegative, bounded and satisfies the zero Neumann conditions at infinity, is usually called  TF.

Ansatz corresponding to (\ref{3-21}) and the reduced  system for
system (\ref{3-20}) have the forms
\begin{eqnarray}\nonumber &&  u=U(\omega), \ \omega=x-\alpha
t, \\ \nonumber &&  v=V(\omega)+\left(\beta t+\gamma
e^t\right)U(\omega)-\gamma e^t,\\ \nonumber &&w=W(\omega)
\end{eqnarray} and
\begin{equation}\label{3-24}\begin{array}{l}U''+\alpha U'+U(1-U)=0,\\
V''+\alpha V'+V(1-U)+U\left(W-\beta\right)=0,\\
dW''+\alpha W'+a_3W(1-W)-a_4UW=0.
 \end{array}\end{equation}

It can be noted that the last equation of system (\ref{3-24}) with
the function $U$ from (\ref{3-22})  has the solutions
\begin{equation}\label{3-26}
W=\frac{1-a_4}{4}\left(1-\tanh\left[\frac{1}{2\sqrt{6}}\,\omega\right]
\right)^2,\end{equation} if $d=a_3=1$ and
\begin{equation}\label{3-26*}
W=1-\frac{1}{4}\left(1-\tanh\left[\frac{1}{2\sqrt{6}}\,\omega\right]
\right)^2,\end{equation} if $d=1, \ a_4=1+a_3.$

Thus, we obtain the solutions of the HGF system (\ref{3-20})  in the
forms
\begin{equation}\label{3-23}\begin{array}{l}  u=\frac{1}{4}\left(1-
\tanh\left[\frac{1}{2\sqrt{6}}\,\omega\right] \right)^2, \ \omega=x-\frac{5}{\sqrt{6}}\,t, \\
v=V(\omega)+\frac{1}{4}\left(\beta t+\gamma e^t\right)\left(1-
\tanh\left[\frac{1}{2\sqrt{6}}\,\omega\right] \right)^2-\gamma e^t,\\
w=\frac{1-a_4}{4}\left(1-\tanh\left[\frac{1}{2\sqrt{6}}\,\omega\right]
\right)^2,
\end{array}\end{equation} if $d=a_3=1$ and \begin{equation}\label{3-23*}\begin{array}{l}  u=\frac{1}{4}\left(1-
\tanh\left[\frac{1}{2\sqrt{6}}\,\omega\right] \right)^2, \ \omega=x-\frac{5}{\sqrt{6}}\,t, \\
v=V(\omega)+\frac{1}{4}\left(\beta t+\gamma e^t\right)\left(1-
\tanh\left[\frac{1}{2\sqrt{6}}\,\omega\right] \right)^2-\gamma e^t,\\
w=1-\frac{1}{4}\left(1-\tanh\left[\frac{1}{2\sqrt{6}}\,\omega\right]
\right)^2
\end{array}\end{equation} if $d=1, \ a_4=1+a_3.$
In (\ref{3-23}) and (\ref{3-23*}),  the function $V$ is an arbitrary
solution of the linear ODE \begin{equation}\label{3-27}V''+\alpha
V'+V(1-U)+U\left(W-\beta\right)=0\end{equation} with  $W$ from
(\ref{3-26}) and (\ref{3-26*}), respectively, while $U$  is given by
formula  (\ref{3-22}).

\begin{remark} Although ODEs (\ref{3-25}) and (\ref{3-27})  are
linear, we were unable to solve them exactly and they are not listed
in the well-known handbooks like \cite{ka-77,po-za-03}.
\end{remark}

Ansatz corresponding to operator (\ref{3-42}) have the form
\begin{equation}\label{3-43}\begin{array}{l}  u=U(t),  \quad
  v=V(t)+\left(\beta t+\gamma e^t\right)x\,U(t)-\gamma e^tx,\quad
w=W(t).
\end{array}\end{equation} Note that the exact solutions of the form (\ref{3-43}) are not important
   from the applicability point of view  because two components ($u$ and $w$)
  depend only on the variable $t$.

\begin{table}
\caption{}\medskip
\label{tab2}       % Give a unique label
\begin{tabular}{p{2cm}p{7cm}p{7cm}}
\hline\hline\noalign{\smallskip}  Operator &Ansatz  & Reduced  system  \\  \hline &&\\
(\ref{3-29}) with \newline{$1+\beta a_1\neq0$}
 & $u=e^{-\beta a_1t}U(\omega)+\frac{\gamma e^t}{1+\beta
 a_1}\left((a_4-1)V(\omega)+\right.$ \newline{$\left.
W(\omega)+(1-a_4)/a_1\right), \ \omega=x-\alpha
t,$}\newline{$v=V(\omega)-\frac{u}{a_1},$}\newline{$w=W(\omega)$}&
$U''+\alpha U'+ U\left(1+a_1\beta-a_1V\right)=0,$
\newline{$V''+\alpha V'+ V\left(1-a_1V+a_1W\right)=0,$}
\newline{$W''+\alpha W'-a_1a_4VW=0$}
  \\ \hline &&\\
(\ref{3-29}) with \newline{$1+\beta a_1=0$}
 & $u=e^{t}\left(U(\omega)+\gamma\left((a_4-1)V(\omega)+\right.\right.$
 \newline{$\left.\left.
W(\omega)+(1-a_4)/a_1\right)t\right), \ \omega=x-\alpha t,$}
\newline{$v=V(\omega)-\frac{u}{a_1},$}\newline{$w=W(\omega)$}
& $U''+\alpha U'-a_1UV-$\newline{$\gamma\left((a_4-1)V
+W+(1-a_4)/a_1\right)=0,$} \newline{$ V''+\alpha V'+
V\left(1-a_1V+a_1W\right)=0,$} \newline{$ W''+\alpha W'-a_1a_4VW=0,$}  \\ \hline &&\\
(\ref{3-29*}) with \newline{$\beta\neq0$}
 & $u=e^{-\beta a_1x}U(t)+\frac{\gamma e^t}{\beta a_1}((a_4-1)V(t)
+$\newline{$W(t)+(1-a_4)/a_1),$}
\newline{$v=V(t)-\frac{u}{a_1},$}\newline{$w=W(t)$}
& $U'+ U\left(a_1V-1-a_1^2\beta^2\right)=0,$ \newline{$ V'+
V\left(a_1V-a_1W-1\right)=0,$}\newline{$ W'+a_1a_4VW=0,$}  \\ \hline &&\\
(\ref{3-29*}) with \newline{$\beta=0$}  &  $u=U(t)+\gamma
e^t((a_4-1)V(t)+$\newline{$W(t)+(1-a_4)/a_1)x,$}
\newline{$v=V(t)-\frac{u}{a_1},$}\newline{$w=W(t)$}
&$U'+ U\left(a_1V-1\right)=0,$\newline{$ V'+
V\left(a_1V-a_1W-1\right)=0,$}
\newline{$ W'+a_1a_4VW=0,$}\\ \hline\hline
\end{tabular}
\end{table}

Finally, we examine the HGF system
\begin{equation}\label{3-28}\begin{array}{l} u_t = u_{xx}+u(1-u-a_1v),\ a_1\neq0,\\ v_t =
 v_{xx}+v(1-u-a_1v)+uw+a_1vw, \\ w_t =  w_{xx}-a_4uw-a_1a_4vw, \end{array}\end{equation}
corresponding to Case 9 of Table~\ref{tab1}. The most general linear
combinations of its  Lie symmetries
\begin{equation}\label{3-29}\begin{array}{l} \hskip1cm
X=\p_t+\alpha\p_x+\beta (-a_1u\p_u+u\p_v)+\\ \gamma
e^t\left(\frac{a_4-1}{a_1}\,u+
(a_4-1)v+w+\frac{1-a_4}{a_1}\right)\left(\p_u-\frac{1}{a_1}\p_v\right)\end{array}\end{equation}
and \begin{equation}\label{3-29*}\begin{array}{l} \hskip1cm
X=\p_x+\beta (-a_1u\p_u+u\p_v)+\\ \gamma
e^t\left(\frac{a_4-1}{a_1}\,u+
(a_4-1)v+w+\frac{1-a_4}{a_1}\right)\left(\p_u-\frac{1}{a_1}\p_v\right)\end{array}\end{equation}
lead to the ans\"{a}tze and the reduced  systems for system
(\ref{3-28}) presented in Table~\ref{tab2}.

\section{Traveling wave solutions and their interpretation}
\label{sec:4}

In this section, we look for TFs (a special subclass of the plane
wave solutions) of the HGF system (\ref{1-2}). TFs are the most
common in theoretical and applied studies of nonlinear real world
models (see, e.g., \cite{britton,mu-89,murr2003,mu-03}). In the case
of a single RD equation, a substantial    number of such solutions
are presented in \cite{gild-ker-04}. Although paper
\cite{ao-sh-shige-96} devoted to study TFs of system (\ref{1-1}),
such solutions are not explicitly presented therein. Here we
construct several TFs of the HGF system (\ref{1-2}) and present
their interpretation.

 As we noted
above, one diffusivity can be set $1$ in (\ref{1-2}) without losing
a generality, hence we consider system
\begin{equation}\label{4-1}\begin{array}{l}  u_t = u_{xx}+u(1-u-a_1v),\\  v_t =
d_2 v_{xx}+ a_2v(1-u-a_1v)+uw+a_1vw,\\  w_t = d_3
w_{xx}+a_3w(1-w)-a_4uw-a_5vw \end{array}\end{equation} in what
follows. Because system (\ref{4-1}) with arbitrary coefficienrs
admits only the trivial algebra (\ref{2-2}),
    the  plane wave ansatz
\[  u=U(\omega), \ \omega=x-\alpha
t, \  v=V(\omega),\  w=W(\omega) \] can be easily derived, which
 reduces  (\ref{4-1}) to the
   nonlinear ODE system
\begin{equation}\label{4-3}\begin{array}{l}U''+\alpha U'+U(1-U-a_1V)=0,\\
d_2V''+\alpha V'+a_2V(1-U-a_1V)+UW+a_1VW=0,\\
d_3W''+\alpha W'+a_3W(1-W)-a_4UW-a_5VW=0.
 \end{array}\end{equation}

Obviously, the ODE system (\ref{4-3}) with arbitrary coefficients is
not integrable, hence, we seek for its particular solutions. Our aim
is to find TFs, i.e.  such plane wave solutions, which are positive
and bounded for arbitrary $x$ and $t>0$.  Moreover, in order to
provide a biological interpretation of determined solutions, we
assume that the solutions to-be-determined connect the steady-state
points of system (\ref{4-1}). Taking into account the arguments
presented above, we consider the ad hoc ansatz
\begin{equation}\label{4-9}\begin{array}{l}
 U= \sigma_1\left(1-\tanh \mu\, \omega \right)^{k_1}, \\
V=\sigma_2\left(1-\tanh \mu\, \omega \right)^{k_2}, \\
W=1-\sigma_3\left(1-\tanh \mu\, \omega \right)^{k_3}.
\end{array}\end{equation}
Notably, ans\"{a}tze of such form are often used and the
corresponding technique is often called the tanh method
\cite{malfliet-04,waz-08}.

\begin{figure}\begin{center}
\begin{minipage}{10cm}
  \quad  \quad \begin{center}\includegraphics[width=8cm]{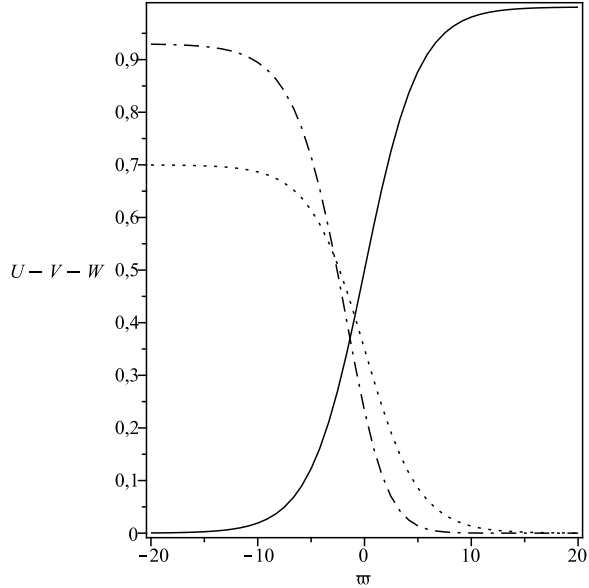}\end{center}
\end{minipage}\end{center}
\caption{Curves representing the functions $U(\omega)$ (dash-dot),
$V(\omega)$ (dot) and $W(\omega)$ (solid)  from
  (\ref{4-4}) for parameters $a_1=1/10$ and $\delta=7/20$.}\label{f2}
\end{figure}

We assume that the exact solution of the form (\ref{4-9}) connects
steady-state points of (\ref{4-3}), namely $(U_0,V_0,0)$ (as $\omega
\rightarrow -\infty$) and $(0,0,1)$ (as $\omega \rightarrow
+\infty$). Having such assumption, one immediately obtains the
restrictions
\[1-2^{k_1}\sigma_1-a_12^{k_2}\sigma_2=0, \ 1-2^{k_3}\sigma_3=0.\]
Substituting ansatz (\ref{4-9}) into (\ref{4-3}) and  making the
corresponding calculations, the exact solution
\begin{equation}\label{4-4}\begin{array}{l}U=\frac{1}{4} (1-2 a_1 \delta ) \left(1-\tanh\left[\frac{
\sqrt{1-2 a_1 \delta }}{2 \sqrt{6}}\,\omega\right]\right)^2, \\
V=\delta -\delta  \tanh\left[\frac{ \sqrt{1-2 a_1 \delta }}{2
\sqrt{6}}\,\omega\right], \\
W=\frac{1}{2} +\frac{1}{2}\tanh\left[\frac{ \sqrt{1-2 a_1 \delta
}}{2 \sqrt{6}}\,\omega\right]
\end{array}\end{equation} of system (\ref{4-3}) was constructed. Here $a_1\leq\frac{1}{2\delta}$ (otherwise the solution is complex),
$\delta>0$ (otherwise  $V$ is negative) and the additional
restrictions \begin{equation}\label{4-5}\begin{array}{l}
\medskip \alpha =\frac{5-4 a_1 \delta }{\sqrt{6-12 a_1 \delta }}, \
d_2=\frac{-3-5 \delta +6 a_1 \delta +4 a_1 \delta ^2}{\delta  (-3+2
a_1 \delta )}, \ a_2=\frac{3-10\delta+6
a_1\delta +8 a_1 \delta ^2}{6 \delta  (-3+2 a_1 \delta)},\\
\hskip2cm
 a_4=\frac{d_3}{3}, \ a_5=\frac{5-d_3+6 a_3-4 a_1 \delta +2
a_1 d_3 \delta }{12 \delta }\end{array}\end{equation} must take
place. Because  $d_2>0, \ a_2\geq0$ and $a_5\geq0$,   the further
restrictions  \begin{equation}\label{4-10}\begin{array}{l}a_3\leq
\frac{1}{6} \left(-5+4 \delta a_1+d_3-2 \delta a_1 d_3\right),\
d_3\geq
\frac{5-4 \delta a_1}{1-2 \delta a_1},\\ a_1\leq \left\{ \begin{array}{l} \frac{1}{2\delta}, \  \mbox{if} \ \delta>1,\\
\frac{-3+10 \delta }{2 \delta (3+4 \delta )}, \
   \mbox{if} \  \frac{3}{10}\leq\delta\leq1
\end{array} \right.
\end{array}\end{equation} are obtain from (\ref{4-5}).

As one can see, the exact solution (\ref{4-4}) is nothing else but
the exact solution
 connecting the steady-state points $\left(1-2 a_1 \delta,2\delta,0\right)$ and
 $(0,0,1)$ of system (\ref{4-3}), because
\begin{eqnarray}\nonumber &&(U,V,W)\rightarrow \left(1-2 a_1 \delta,2\delta,0\right) \
 \mbox{if} \ \omega \rightarrow -\infty, \\ \nonumber && (U,V,W)\rightarrow
(0,0,1) \  \mbox{if} \ \omega \rightarrow +\infty.
\end{eqnarray} An example of the exact solution
(\ref{4-4}) is presented in Fig.\,\ref{f2}.

\begin{figure}\begin{center}
\begin{minipage}{10cm}
  \quad  \quad \begin{center}\includegraphics[width=8cm]{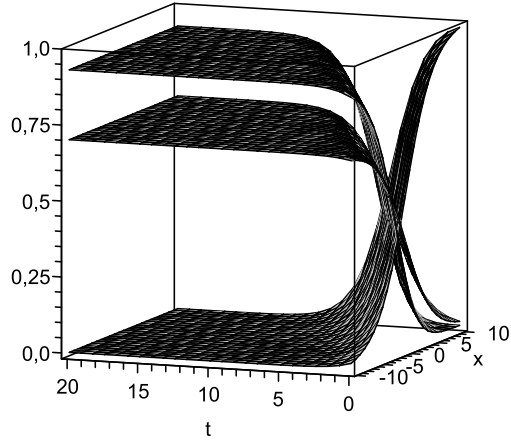}\end{center}
\end{minipage}\end{center} \caption{Surfaces representing  TF
(\ref{4-11}) with $\alpha=81/5\sqrt{62}$ and $\delta=7/20$ of the
HGF system (\ref{4-1}) with the parameters $a_1=1/10, \ a_2=64/2051,
\ a_4=d_3/3, \ a_5=(162+200a_3-31d_3)/140, \ d_2=8982/2051$.
}\label{f1}
\end{figure}

Thus, the one-parameter family of TFs
\begin{equation}\label{4-11}\begin{array}{l}u=\frac{1}{4} (1-2 a_1 \delta ) \left(1-\tanh\left[\frac{
\sqrt{1-2 a_1 \delta }}{2 \sqrt{6}}(x-\alpha t)\right]\right)^2, \\
v=\delta -\delta  \tanh\left[\frac{ \sqrt{1-2 a_1 \delta }}{2
\sqrt{6}}(x-\alpha t)\right], \\
w=\frac{1}{2} +\frac{1}{2}\tanh\left[\frac{ \sqrt{1-2 a_1 \delta
}}{2 \sqrt{6}}(x-\alpha t)\right]
\end{array}\end{equation} of the HGF system (\ref{4-1}) with
restrictions (\ref{4-5})--(\ref{4-10}) is derived. This solution has
a clear biological interpretation and describes such interaction
between  farmers and hunter-gatherers that hunter-gatherers die
while the initial and converted farmers coexist (see
Fig.\,\ref{f1}). Actually, one may say that extinction of
hunter-gatherers takes place because all of them are converted into
farmers.

Now we turn to  system (\ref{4-1}) with $a_1=0$:
\begin{equation}\label{4-6}\begin{array}{l}  u_t = u_{xx}+u(1-u),\\  v_t =
d_2 v_{xx}+ a_2v(1-u)+uw,\\  w_t = d_3 w_{xx}+a_3w(1-w)-a_4uw-a_5vw.
\end{array}\end{equation}
It follows from Theorem~\ref{th-1} that system (\ref{4-6}) with
$a_2\neq0$ and $a_3+a_5>0$ admits only the trivial algebra
(\ref{2-2}).

In order  to construct exact solution of system (\ref{4-6}), an
analog of ad hoc ansatz (\ref{4-9}) and additional restrictions have
been again used. As a result, TF
\begin{equation}\label{4-8}\begin{array}{l} u(t,x)\equiv U(\omega) =
\frac{1}{4}\left(1-\tanh\left[\frac{1}{2\sqrt{6}}\,\omega\right]
\right)^2, \ \omega=x-
\frac{5}{\sqrt{6}}\,t,\\
v(t,x)\equiv V(\omega) =
\frac{3d-5}{3(d-5)}\left(1-\tanh\left[\frac{1}{2\sqrt{6}}\,\omega\right]
\right)^3,  \\  w(t,x)\equiv W(\omega) =
\frac{3d-5}{2(d-5)}\left(1-\tanh^2\left[\frac{1}{2\sqrt{6}}\,\omega\right]
\right)
\end{array}\end{equation}
of the system
\begin{eqnarray}\nonumber && u_t = u_{xx}+u(1-u),\\ \nonumber && v_t =
\frac{1}{2}\, v_{xx}+ v(1-u)+uw,\\ \nonumber && w_t = d
w_{xx}+\frac{5-d}{6}\,w(1-w)-\frac{5}{3}\,uw
\end{eqnarray} was constructed (here $d\leq5/3$).

TF  (\ref{4-8})
 connects the steady-state points  $\left(1,\frac{8(3d-5)}{3(d-5)},0\right)$ and
 $(0,0,0)$ because
\begin{eqnarray}\nonumber && (U,V,W)\rightarrow \left(1,\frac{8(3d-5)}{3(d-5)},0\right) \
 \mbox{if} \ \omega \rightarrow -\infty, \\ \nonumber && (U,V,W)\rightarrow
(0,0,0) \  \mbox{if} \ \omega \rightarrow +\infty.
\end{eqnarray} Thus, the biological interpretation of
solution (\ref{4-8}) is similar to that for solution (\ref{4-11}).
Notably, TF  (\ref{4-8}) in contrast to that  (\ref{4-11}) has the
fixed wave velocity $\alpha=\frac{5}{\sqrt{6}},$ which is exactly
the same as for TF (\ref{3-22}) of the Fisher equation.

\begin{remark} Because the HGF system (\ref{1-2}) is invariant with
respect to the discrete transformation $x \rightarrow -x$, all the
solutions obtained above can be transformed to another solutions
using this transformation.
\end{remark}

\section{Conclusions} \label{sec:5}

 In this paper,  the three-component nonlinear system of PDEs (\ref{1-1}) introduced in
  \cite{ao-sh-shige-96} for describing the spread of an initially
   localized population of farmers into a  region occupied by
   hunter-gatherers was studied by the classical Lie method. First of all the system was transformed to
    the nondimensional form (\ref{1-1*}) in order to reduce the number of parameters.
    All possible Lie symmetries of  system (\ref{1-1*}) were identified
    (Theorem\ref{th-1}), inequivalent  symmetry reductions to
    the ODE systems   in the most interesting case
     (from applicability point of view) were conducted (Section~\ref{sec:3}),
     several families of exact solutions (including the travelling fronts) were found and
       a possible biological  interpretation for  some of them was provided (Section~\ref{sec:4}).

  It is worth noting that the nonlinear  system (\ref{1-1}) was studied under the restriction $e_1\not=0$,
  otherwise the system reduces to the three-component  diffusive Lotka--Volterra system (DLVS).
 Lie symmetries of the three-component  DLVS are completely described in
\cite{che-dav2013}, while its exact solutions were constructed in
\cite{che-dav2013, chen-hung-12,hung-11}.

 To the best of our knowledge, this paper  is the first study of the
 HGF model by symmetry-based methods.
 In \cite{ao-sh-shige-96}, the authors  studied the existence and behaviour of TFs of the model, however any  exact solutions are not presented therein. In particular, it is stated that there are TFs connecting the stable and unstable steady-state points of the model (see P.10 in \cite{ao-sh-shige-96}). Interestingly  that TF (\ref{4-11}) corresponds exactly to  such case provided restrictions (\ref{4-5})--(\ref{4-10}) hold. Moreover, we constructed the exact solution (\ref{3-41}), which predicts coexistence of all the populations at any semifinal space interval (see formulae (\ref{3-41**})) provided the coefficients of the HGF system (\ref{1-2}) satisfy the
 restrictions (\ref{3-41*}). Such type of behaviour was not identified in \cite{ao-sh-shige-96}.

 A natural continuation of this research is searching for non-Lie (nonclassical, conditional, etc.)  symmetries   of the nonlinear  system (\ref{1-1})  and their application for constructing exact solutions. We have achieved  some  progress  in this direction and plan to report new results in a forthcoming paper.

\end{document}